\documentclass[conference]{IEEEtran}
\IEEEoverridecommandlockouts
\usepackage{cite}
\usepackage{amsmath,amssymb,amsfonts}
\usepackage{algorithmic}
\usepackage{booktabs}
\usepackage{graphicx}
\usepackage{textcomp}
\usepackage{xcolor}
\usepackage{multirow}
\usepackage{algorithm}
\usepackage{multirow}
\usepackage{amsmath}
\usepackage{hyperref}
\usepackage{balance}
\usepackage{subfigure}
\usepackage{threeparttable}

\def\BibTeX{{\rm B\kern-.05em{\sc i\kern-.025em b}\kern-.08em
    T\kern-.1667em\lower.7ex\hbox{E}\kern-.125emX}}

\begin{document}

\title{ForgeEDA: A Comprehensive Multimodal Dataset for Advancing EDA}

\author{
	\IEEEauthorblockN{
		Zhengyuan Shi$^{1, 9}$,
        Zeju Li$^{1, 9}$,
		Chengyu Ma$^{2, 9}$, 
        Yunhao Zhou$^{1, 9}$, 
		Ziyang Zheng$^{1, 9}$,
        Jiawei Liu$^{3, 9}$,
        Hongyang Pan$^{4, 9}$, \\
		Lingfeng Zhou$^{5, 9}$,
        Kezhi Li$^{1, 9}$,
        Jiaying Zhu$^{1, 9}$,
        Lingwei Yan$^{6, 9}$, 
        Zhiqiang He$^{6, 9}$,
        Chenhao Xue$^{7, 9}$,
        Wentao Jiang$^{2, 9}$, \\
        Fan Yang$^4$,
        Guangyu Sun$^7$,
        Xiaoyan Yang$^5$,
        Gang Chen$^6$,
        Chuan Shi$^3$,
        Zhufei Chu$^2$,
        Jun Yang$^{8,9}$
        and Qiang Xu$^{1,9*}$\thanks{This work was partly supported by the General Research Fund of the Hong Kong Research Grants Council (RGC) (Grant Nos. 14212422 and 14202824), National Technology Innovation Center for EDA, National Natural Science Foundation of China (Grant Nos. 62032001, U20B2045, U1936220, 62192784, 62172052, 62002029, 61772082, 92373207), Beijing Natural Science Foundation (Grant No. L243001), and the 111 Project (Grant No. B18001).}} 

\IEEEauthorblockA{$^1$\textit{Department of Computer Science and Engineering}, \textit{The Chinese University of Hong Kong}, Sha Tin, Hong Kong S.A.R.\\}
\IEEEauthorblockA{$^2$\textit{Faculty of Electrical Engineering and Computer Science}, \textit{Ningbo University}, Ningbo, China \\}
\IEEEauthorblockA{$^3$\textit{School of Computer Science}, \textit{Beijing University of Posts and Telecommunications}, Beijing, China\\}
\IEEEauthorblockA{$^4$\textit{School of Microelectronics, State Key Laboratory of Integrated Chips and System}, \textit{Fudan University}, Shanghai, China\\}
\IEEEauthorblockA{$^5$\textit{School of Electronic and Information Engineering}, \textit{Hangzhou Dianzi University}, Hangzhou, China \\}
\IEEEauthorblockA{$^6$\textit{College of Computer Science and Technology}, \textit{Nanjing University of Aeronautics and Astronautics}, Nanjing, China\\}
\IEEEauthorblockA{$^7$\textit{School of Integrated Circuits}, \textit{Peking University}, Beijing, China \\}
\IEEEauthorblockA{$^8$\textit{School of Intergrated Circuits}, \textit{Southeast University}, Nanjing, China \\}
\IEEEauthorblockA{$^9$\textit{National Center of Technology Innovation for EDA}, Nanjing, China \\}
\IEEEauthorblockA{Corresponding Author: Qiang Xu (qxu@cse.cuhk.edu.hk) \\}
} 

\maketitle

 % \begingroup\renewcommand\thefootnote{\textsection}
 % \footnotetext{Both authors contributed equally to this research.}
 % \endgroup

\begin{abstract}
We introduce ForgeEDA, an open-source comprehensive circuit dataset across various categories. ForgeEDA includes diverse circuit representations such as Register Transfer Level (RTL) code, Post-mapping (PM) netlists, And-Inverter Graphs (AIGs), and placed netlists, enabling comprehensive analysis and development.  We demonstrate ForgeEDA's utility by benchmarking state-of-the-art EDA algorithms on critical tasks such as Power, Performance, and Area (PPA) optimization, highlighting its ability to expose performance gaps and drive advancements. Additionally, ForgeEDA's scale and diversity facilitate the training of AI models for EDA tasks, demonstrating its potential to improve model performance and generalization. By addressing limitations in existing datasets, ForgeEDA aims to catalyze breakthroughs in modern IC design and support the next generation of innovations in EDA. 
\end{abstract}


\section{Introduction} \label{Sec:Intro}

% Modern chip design is a multifaceted endeavor characterized by its intricacy, involving a diverse array of sub-modules, alongside numerous stages within the Electronic Design Automation (EDA) flow. The process of designing typically entails collaboration across many departments in a semiconductor company to translate the design specification into Register Transfer Level (RTL) code, netlist, floorplanning and layout. 

% To enhance the existing EDA tools and speed time-to-market, integration of Artificial Intelligence (AI) technique in EDA has emerged as an attractive direction. For instance, DeepGate Family~\cite{li2022deepgate, shi2023deepgate2, shi2024deepgate3} are trained on a large number of circuit netlists and employed into design for test~\cite{shi2022deeptpi} and power analysis~\cite{khan2024deepseq}. Besides, ~\cite{allam2024RTL-Repo, cui2024OriGen, thakur2023benchmarking} learn web-scale code and benefit hardware code generation. Clearly, the effectiveness of deep learning models is heavily contingent upon the quality of the training data~\cite{chang2024data}. However, we observe that the existing circuit datasets in EDA field suffer from notable limitations.

Electronic Design Automation (EDA) is an indispensable technology in the semiconductor industry, breaking down the complexity of integrated circuit (IC) design into a multi-stage process, including logic synthesis, technology mapping, placement, routing, and layout generation. Over the past several decades, researchers in the EDA field have worked to address the challenges in reducing design costs, enhancing chip PPA metrics, and ensuring design functionality. Thanks to the advanced EDA solutions~\cite{wang2009electronic} and the emerging AI for EDA (AI4EDA) solutions~\cite{huang2021machine, chen2024large}, engineers are now capable of designing large-scale and diverse ICs, from billion-transistor processors and AI accelerators to compact low-power chips. 

However, despite these achievements, the hardware design ecosystem remains highly closed compared to software community. The lack of accessible data has been a significant barrier to evaluating novel EDA technologies and training AI4EDA models, which in turn has hindered the rapid progress of the entire hardware design. Existing datasets~\cite{ISCAS89, ITC99, EPFLBenchmarks, chowdhury2021openabc, chai2022circuitnet, jiang2024circuitnet2} offer open-access circuits but suffer from limitations in size and variety. For example, the largest circuit in ITC99 benchmark~\cite{ITC99} contains only ten thousands of gates and CircuitNet~\cite{chai2022circuitnet, jiang2024circuitnet2} only includes processors, making them insufficient for addressing the needs of modern IC design. 
Additionally, datasets like~\cite{thakur2023benchmarking} collect web-scale hardware code snippets for training large language models (LLMs). While valuable for specific applications such as RTL understanding and generation, they do not support the remaining EDA stages and are thus limited to narrow use cases. These shortcomings highlight a critical gap in existing EDA datasets, underscoring the need for a comprehensive, practical, and large-scale dataset to support the next generation of EDA advancements. 

% We categorize the existing circuit datasets into three groups and observe the notable limitations. 
% First, ~\cite{} utilize proprietary industrial data to assess the effectiveness of EDA algorithms. However, these datasets are not freely accessible, limiting their use in the broader research community. 
% Second, other datasets~\cite{ISCAS89, ITC99, EPFLBenchmarks, chowdhury2021openabc, chai2022circuitnet, jiang2024circuitnet2} offer open access but suffer from limitations in size and variety. For example, the largest circuit in ITC99 benchmark~\cite{ITC99} contains only ten thousands of gates and CircuitNet~\cite{chai2022circuitnet, jiang2024circuitnet2} only includes processors, making them insufficient for addressing the needs of modern IC design. 
% Third, datasets like~\cite{thakur2023benchmarking} collect web-scale hardware code snippets for training large language models (LLMs). While valuable for specific applications such as RTL understanding and generation, they do not support downstream circuit representations and are thus limited to narrow use cases. 

\begin{figure}[!t]
    \centering
    \includegraphics[width=1.0\linewidth]{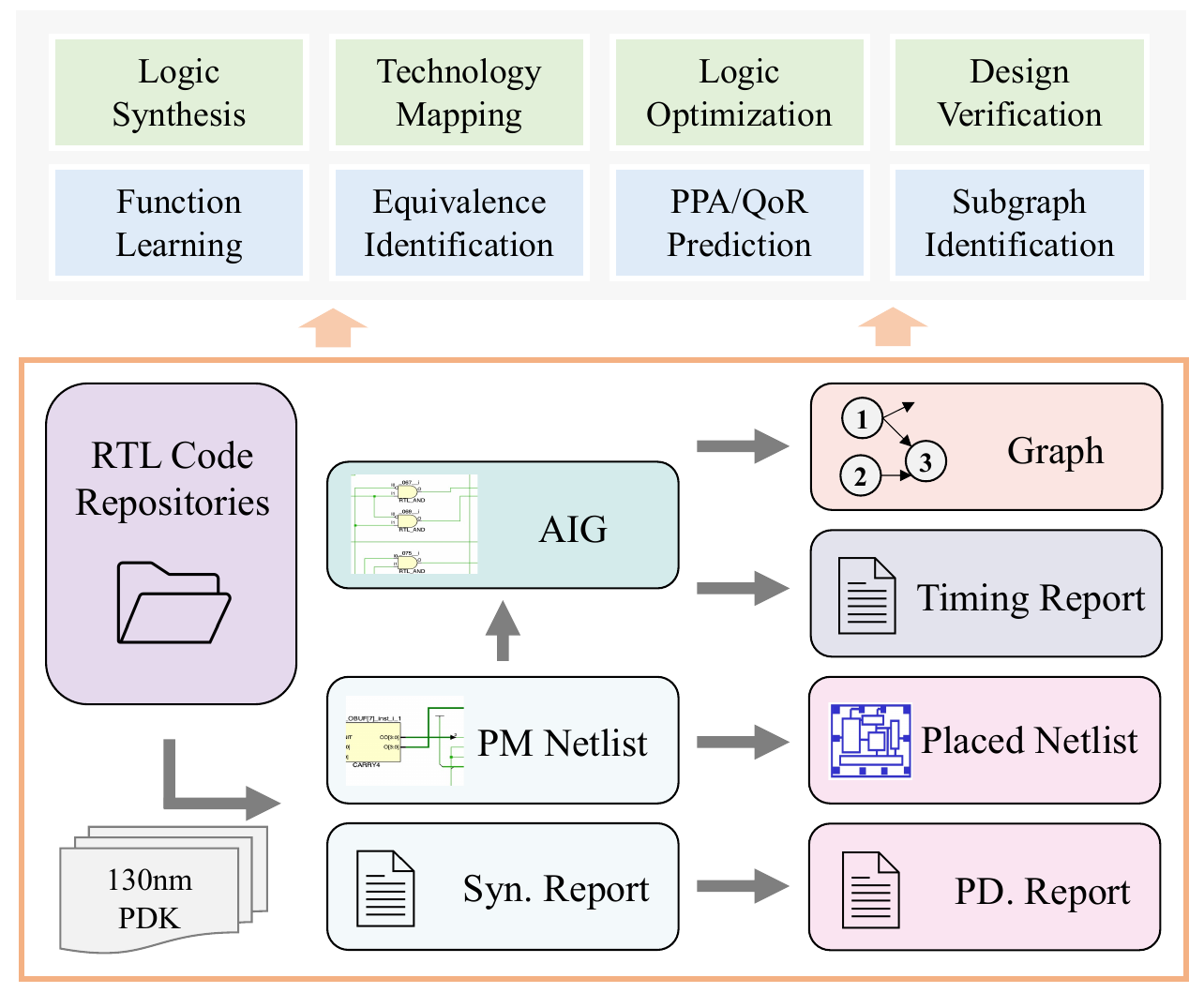}
    \caption{Overview of ForgeEDA}
    \label{fig:overview}
\end{figure}

To address the above issues, we propose ForgeEDA, a comprehensive multimodal dataset designed to advance EDA research. ForgeEDA spans various circuit formats across the EDA flow and encompasses a wide range of real-world chip types. Specifically, the dataset includes 6 categories, 20 divisions, and 1,189 repositories that reflect real-world circuit designs, such as RISC-V cores, AI accelerators, arithmetic units, encoders/decoders, interface modules, and control units (see Table.~\ref{Tab:types}). In addition to the original Verilog code, ForgeEDA provides the corresponding Post-Mapping (PM) netlists, placed netlists, and And-Inverter Graphs (AIGs), generated using EDA tools.

We utilize ForgeEDA to assess EDA solutions, focusing on logic synthesize and optimization. Our evaluation reveals a significant performance gap between open-source EDA solutions and advanced commercial tools, a disparity that is less evident in previous datasets due to their narrow scopes or scales. Additionally, we leverage our dataset for training AI4EDA solutions. The expanded scale and diversity of training data enable our dataset to enhance existing AI4EDA models, driving further improvements in their performance. 
% Finally, we explore the potential future applications of ForgeEDA. 
We believe that our dataset can pave the way for advancing the next generation of EDA.

\begin{table}[!t]
\caption{Categories of Collected Designs}  \label{Tab:types}
\centering
\renewcommand\tabcolsep{3.0pt}
\begin{tabular}{@{}cllcll@{}}
\toprule
\multicolumn{1}{l}{Category}       & Division          & \multicolumn{1}{l|}{Counts} & \multicolumn{1}{l}{Category} & Division          & Counts        \\ \midrule
\multirow{4}{*}{Processor}         & RISCV             & \multicolumn{1}{l|}{111}    & \multirow{4}{*}{Interface}   & SPI               & 45            \\
                                   & MIPS              & \multicolumn{1}{l|}{26}     &                              & UART              & 40            \\
                                   & AI                & \multicolumn{1}{l|}{53}     &                              & PCI               & 30            \\
                                   & Others  & \multicolumn{1}{l|}{58}     &                              & Others  & 32            \\ \midrule
\multirow{5}{*}{arithmetic}        & Adder             & \multicolumn{1}{l|}{38}     & \multirow{5}{*}{Controller}  & DMA               & 30            \\
                                   & Multipiler        & \multicolumn{1}{l|}{109}    &                              & PWM               & 94            \\
                                   & Shifter           & \multicolumn{1}{l|}{15}     &                              & Arbiter           & 43            \\
                                   & Counter           & \multicolumn{1}{l|}{66}     &                              & Clock Control     & 31            \\
                                   & Others & \multicolumn{1}{l|}{66}     &                              & Others & 35            \\ \midrule
\multicolumn{2}{c}{Encoder/Decoder}                    & \multicolumn{1}{l|}{182}    & \multicolumn{2}{c}{Others}                       & 85            \\ \midrule
\multicolumn{1}{l}{\textbf{Total}} &                   &                             & \multicolumn{1}{l}{}         &                   & \textbf{1,189} \\ \bottomrule
\end{tabular}
\end{table}

\section{Related Work} \label{Sec:Related}
\subsection{Previous Dataset in EDA}
A comprehensive and unified circuit dataset is essential for the fair evaluation of EDA solutions and serve as an infrastructure for training AI4EDA models. Some existing datasets focus on specific circuit formats tailored for a single task, like the EPFL benchmark~\cite{EPFLBenchmarks} for logic synthesis or a dataset comprising collected RTL code snippets~\cite{thakur2023benchmarking} utilized to fine-tune large language models (LLMs) for Verilog code generation and comprehension. Additionally, prominent organizations host contests or annual competitions and release corresponding benchmarks such as ISCAS89~\cite{ISCAS89}, ITC99~\cite{ITC99}, HWMCC~\cite{biere2024hardware}, and SAT competitions~\cite{gbd}. However, these datasets are limited in the types and scales of circuits and are often constrained to single-task evaluations. Other datasets provide multiple circuit representations. For instance, OpenABC-D~\cite{chowdhury2021openabc} presents a comprehensive circuit dataset for logic synthesis, encompassing RTL code, circuit netlists, and graph-based representations, whereas the CircuitNet~\cite{chai2022circuitnet, jiang2024circuitnet2} extends the EDA process to the physical design phase by including floorplans and layouts. Nevertheless, these datasets have constraints in their coverage. OpenABC-D, for example, is limited to only 29 open-source hardware IP designs, while CircuitNet primarily focuses on processors. Different from the previous datasets, our ForgeEDA covers 7 categories and 22 sub-categories of circuit designs, providing RTL code, netlists and AIGs. 

\subsection{EDA Applications}
\subsubsection{Logic Synthesis}
Logic synthesis is a cornerstone of EDA design flow, enabling the transformation of high-level hardware description languages (HDL) into optimized netlists of logic gates~\cite{hassoun2001logic}. Traditional logic synthesis focuses on balancing design constraints and meeting design objectives such as minimizing area, power consumption, and delay~\cite{mishchenko2011scalable, mishchenko2006dag}. Over time, advances in this field have introduced techniques for fine-tuning synthesis processes to better align with downstream physical design stages, improving efficiency throughout the design flow~\cite{zhu2023delay, pan2024physically}. However, the de-facto pipeline for evaluating logic synthesis methodologies is heavily reliant on narrow and small-scale benchmarks~\cite{ISCAS89, EPFLBenchmarks}. Performance on these benchmarks has nearly converged, making it difficult to assess the generalization capability of synthesis approaches across more diverse and complex scenarios. We leverage our ForgeEDA dataset to evaluate a range of logic synthesis methods, providing a more rigorous and diverse testing ground that mirrors real-world design challenges. 
% By comparing these methods against state-of-the-art commercial tools, we reveal limitations in current academic synthesis tools, which often struggle to match the robustness and adaptability of commercial solutions. 

\subsubsection{PPA Prediction}
PPA prediction has emerged as a critical focus in early-stage IC design due to its potential to streamline design cycles by enabling informed decision-making. Recent research has focused on developing pre-synthesis PPA estimation frameworks. For example, MasterRTL~\cite{fang2023masterrtl} introduces a bit-level design representation, the Simple Operator Graph (SOG), which is readily available from RTL code and enables more precise PPA predictions. Another notable framework, SNS~\cite{xu2022sns}, leverages a path-based approach by employing deep learning models to estimate PPA at the path level. The existing approaches are typically evaluated on a limited range of circuit types and focus on PPA metrics at post-synthesis stages, lacking the capacity to account for the full physical design process. The physical design stage is both time-intensive and critical for accurate PPA prediction, as it reflects a more realistic measure of the final design's performance and constraints. Currently, no benchmark provides an integrated dataset encompassing RTL code, netlists, and corresponding physical design reports. 

\subsubsection{AI for EDA}
The integration of AI into EDA has opened new avenues for automating and optimizing various design processes~\cite{huang2021machine, chen2024large}. With the advanced capabilities of AI models, AI4EDA solutions have demonstrated remarkable achievements in circuit optimization~\cite{chowdhuryretrieval, song2024circuitvae, shi2022deeptpi}, design metric prediction~\cite{liu2016efficient, ustun2020accurate} and hardware code generation~\cite{liu2024rtlcoder}. Besides, an emerging area within AI4EDA is circuit representation learning~\cite{li2022deepgate, shi2023deepgate2, khan2024deepseq, zheng2025deepgate4}, which learns general representations of circuit designs that can be easily transferred across various EDA tasks. Moreover, DeepGate3~\cite{shi2024deepgate3} proves the data scaling ability of circuit representation learning models, indicating that the model performance can be further improved with more training data. Clearly, a critical prerequisite for the success of AI4EDA solutions is access to high-quality, diverse datasets that encompass a wide range of circuit types and design stages. In this work, we train a series of models with ForgeEDA to demonstrate its utility and effectiveness.

\begin{table*}[!t]
\caption{Contents of ForgeEDA Dataset} \label{TAB:content}
\vspace{-5pt}
\centering
\begin{tabular}{@{}llll@{}}
\toprule
Modality                        & \#Samples               & Format & Contents                                                \\ \midrule
Code Repository                 & 1,189                   & .v     & Verilog code with annotated comments and specifications \\ \midrule
\multirow{2}{*}{PM Netlist}     & \multirow{2}{*}{4,450}  & .v     & Post-mapping netlist                                    \\
                                &                         & .rpt   & Reports of area, path delay and power                   \\ \midrule
\multirow{2}{*}{Placed Netlist} & \multirow{2}{*}{4,450}  & .v / .def     & Placed netlist                                          \\
                                &                         & .rpt   & Reports of area, WNS, TNS and power                     \\ \midrule
AIG                             & 4,450                   & .aig   & Raw AIG files                                           \\ \midrule
\multirow{2}{*}{Sub-AIG}        & \multirow{2}{*}{83,155} & .aig   & Raw AIG files                                           \\
                                &                         & .npz   & Parsed graphs in PyTorch Geometric                                    \\ \bottomrule
\end{tabular}
\vspace{-5pt}
\end{table*}
\section{Dataset} \label{Sec:Dataset}
\subsection{Overview}
The overview of our proposed dataset is illustrated in Fig~\ref{fig:overview}. To construct the dataset, we first collect 1,189 RTL code repositories from the Internet. The categories of the designs in our dataset are summarized in Table~\ref{Tab:types}. Next, we employ EDA tools to generate various circuit formats, including post-mapping (PM) netlists and And-Inverter Graphs (AIGs), accompanied by logs detailing PPA metrics. All the contents of our dataset are summarized in Table~\ref{TAB:content}. Finally, we evaluate our dataset in both practical EDA applications and AI for EDA tasks.

\subsection{Data Preparation}
\subsubsection{Data Collection}
We begin by identifying and extracting design-related keywords from chip specifications available on AllDatasheet~\cite{alldatasheet}, a website hosting millions of semiconductor datasheets. This approach ensures that our dataset covers a diverse array of design types, reflecting real-world chip applications. Next, we gather repositories in Verilog (.v) from the Internet based on these selected keywords to build the RTL code dataset~\cite{li2025deepcircuitx,liu2025deeprtl}. Finally, we filter out repositories that cannot be synthesized or contain only post-mapping Verilog files, refining our dataset to include only those with comprehensive and synthesizable designs. We totally collect 1,189 high-quality open-source repositories in Verilog and categorize them into 6 fields. 

\subsubsection{Logic Synthesis and Physical Design}
We employ Synopsys Design Compiler and Skywater 130nm~\cite{edwards2021introduction} Process Design Kit (PDK) to synthesize RTL code repositories into post-mapped (PM) netlists, along with detailed synthesis reports. To expand our dataset, each module is treated as the top module in its respective synthesis flow. As a result, a total of 4,450 netlists are generated. 
Following synthesis, we utilize Cadence Innovus to carry out the floorplanning and placement steps in the ASIC physical design flow. This generates a placed netlist and a physical design report. After placement, the precise timing information can be extracted since the distances between ports and cells are established.

\subsubsection{Graph Generation}
We begin by converting PM netlists, based on the standard cell library in the PDK, into And-Inverter Graphs (AIGs) using the ABC tool \cite{mishchenko2007abc}. Next, we apply the Static Timing Analysis (STA) tool, \textit{stime}, within ABC to generate the timing report. To support the AI4EDA solution, we construct a dataset using graphs represented in PyTorch Geometric \cite{fey2019fast}. These graphs are derived from the PM netlists and AIGs. We also provide the sub-AIGs for model training, which are randomly extracted sub-circuits with 500-5,000 nodes. Totally, 83,155 sub-circuits and their graph representations are generated for model training.

% To enable the AI4EDA solution, we also build a dataset for model training by parsing the AIGs into graphs in Pytorch Geometric and providing supervision labels, such as logic-1 probability under random simulation and the pairwise similarity~\cite{shi2023deepgate2}. 

\section{Practical EDA Applications} \label{Sec:TradApp}

\begin{table}[!t]
\caption{RTL Synthesis Results} \label{TAB:LS}
\vspace{-5pt}
\renewcommand\tabcolsep{2.0pt}
\begin{tabular}{@{}cl|cc|cc@{}}
\toprule
\multicolumn{2}{c|}{\multirow{2}{*}{Circuit}} & \multicolumn{2}{c|}{DCU}                         & \multicolumn{2}{c}{Yosys}                                     \\
\multicolumn{2}{c|}{}                         & area ($\mu m^2$)                         & delay($ps$)                     & area ($\mu m^2$)                    & delay($ps$)                     \\ \midrule
\multirow{4}{*}{Processor}    & riscv1        & 189,973.45                     & 10,596.26                      & 195,852.83                    & 36,679.18                     \\
                              & riscv2        & 122,628.86                     & 63,897.15                      & 764,719.69                    & 7,772.28                      \\
                              & riscv3        & 20,111.79                      & 17,155.09                      & 41,422.23                     & 10,554.09                     \\
                              & imageProc     & 552,800.19                    & 16,575.54                     & 1,765,114.12                  & 182,101.42                    \\ \midrule
\multirow{4}{*}{Arithmetic}   & alu1          & 3,160.53                      & 6,197.87                      & -                             & -                             \\
                              & alu2          & 1,631.56                      & 10,570.08                     & 41,684.98                     & 140,038.53                    \\
                              & alu3          & 36,488.75                      & 138,830.16                     & 43,481.70                     & 130,999.24                    \\
                              & alu4          & 17,027.58                      & 14,461.29                      & 17,516.80                     & 6,059.55                      \\ \midrule
\multirow{4}{*}{En/Decoder}   & qcLdpc        & 3,589,661.50                  & 42,341.13                     & -                             & -                             \\
                              & ldpc          & 368,947.05                        & 8,386.78                       & 235,082.95                    & 6,011.34                      \\
                              & viterbi       & 147,456.42                     & 82,473.52                      & -                              &  -                             \\
                              & dct           & 49,444.92                     & 14,142.23                     & 72,768.54                     & 18,663.73                     \\ \midrule
\multirow{4}{*}{Interface}    & pcie1         & 118,623.77                     & 10,276.88                      & 140,099.38                    & 8,975.54                      \\
                              & pcie2         & 53,663.97                      & 9,011.58                       & -                             & -                             \\
                              & pcie3         & 61,327.57                      & 105,812.45                     & 94,431.82                     & 33,475.29                     \\
                              & axis          & 4,569.38                       & 12,909.13                      & 3,687.29                      & 4,893.00                      \\ \midrule
\multirow{4}{*}{Controller}   & memctrl    & 181,391.47                     & 10,191.19                      & 432,793.81                    & 2,774.81                      \\
                              & branch        & 41,039.36                      & 3,927.65                       & 51,365.51                     & 129,410.59                    \\
                              & signalTab   & 15,531.15                      & 2,838.90                       & 18,611.60                     & 3,370.16                      \\
                              & arbiter       & 3,168.04                      & 1,291.89                      & {4,332.91}  & 1,435.12                      \\ \midrule
\multicolumn{1}{l}{Geomean}   &               & {48,523.05} & {14,604.15} & {71,987.59} & {15,944.40} \\
\multicolumn{1}{l}{\textbf{Imp.}}      &               & \textbf{1.00}                          & \textbf{1.00}                          & \textbf{1.77}      & \textbf{1.20}      \\ \bottomrule
\end{tabular}
% \begin{tablenotes}
%     \footnotesize
%     \item[1] {\bf *}: Data with this superscript derived only from successful processing (excluding - data).
% \end{tablenotes}
\vspace{-5pt}
\end{table}

To validate the effectiveness of our dataset, we perform a comprehensive evaluation to measure the performance gap between open-source and commercial synthesis tools.
For this assessment, we employed a variety of advanced open-source tools and commercial EDA solutions across the following three main logic synthesis tasks.

\subsection{RTL Synthesis} 
We performed a complete synthesis flow starting from RTL input, which included logic optimization and technology mapping, to generate gate-level netlists. 
We then evaluated and compared the area and delay metrics of these netlists. 
Specifically, we analyzed the performance of the commercial tool Design Compiler's \texttt{compile\_ultra} command~(DCU)~\cite{24_dc} and the open-source logic synthesis tool Yosys~\cite{16_Wolf_yosys}. Table~\ref{TAB:LS} summarizes the post-synthesis delay and area metrics for both methods, which is reported by STA command \texttt{stime} in ABC~\cite{mishchenko2007abc}. 
Out of 20 benchmarks, Yosys failed to produce results for 4 cases due to limitations in its verilog parser, a common drawback in many academic tools. 
For the 16 benchmarks that Yosys successfully processed, the tool achieved an average delay of 1.20$\times$ and an area of 1.77$\times$ compared to DCU. 
These results highlight the significant scalability advantage of commercial tools like DCU over academic tools, while also suggesting that academic tools, such as Yosys, tend to perform better in delay optimization than in area optimization.

\subsection{AIG Optimization} 
To evaluate the capabilities of academic tools in AIG synthesis, we focused on both AIG optimization and AIG technology mapping. We use the selected circuits in Table~\ref{TAB:LS} in the following experiments. 
AIG optimization, which aims to convert an initial AIG into a more optimized form, is assessed with area and delay metrics. We approximate area reduction using the number of nodes reduced and delay reduction using the logic level reduced. 

To comprehensively assess the performance of these methods, we compared the following four approaches:
\begin{itemize}
    \item \texttt{resyn2rs}/\texttt{compress2rs}: Script focuses on minimizing AIG nodes without increasing the logic depth.
    \item \texttt{if -g}: Logic refactoring algorithm that reduces network depth by utilizing sum-of-product~(SOP) expressions~\cite{11ICCAD_mishchenko_delay}.
    \item \texttt{orchestrate}: Combines multiple operations into a single traversal of the AIG for enhanced coordination and optimization~\cite{24TCAD_Li_DAG}.
\end{itemize}

\begin{figure}[!t]
    \centering
    \includegraphics[width=0.9\linewidth]{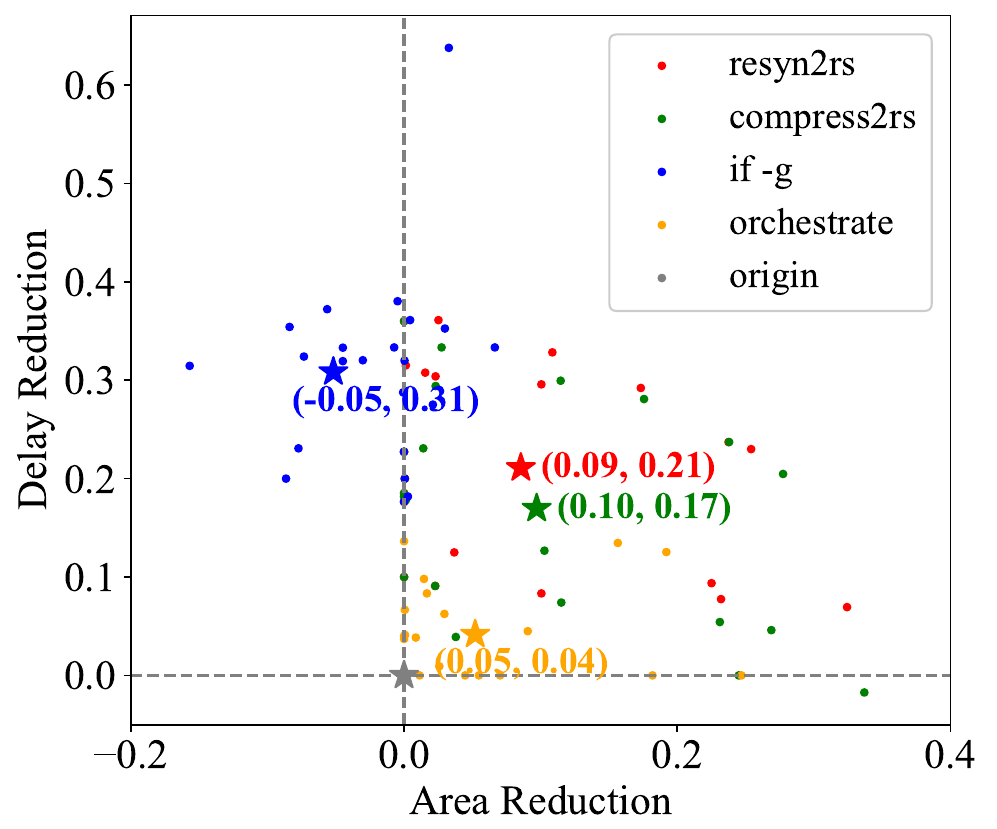}
    \vspace{-5pt}
    \caption{AIG Optimization Results}
    \label{fig:aigOpt}
    \vspace{-10pt}
\end{figure}

In Fig.~\ref{fig:aigOpt}, each node corresponds to an AIG optimized by a specific approach, with its coordinates representing area reduction and delay reduction. The origin of the coordinate system represents the original, unoptimized AIG. The geometric mean of optimization performance across all benchmarks is depicted as `$\star$'. The results reveal several key observations. First, \texttt{resyn2rs} and \texttt{compress2rs} demonstrate the most balanced optimization performance, achieving significant improvements in both area and delay across benchmarks. Second, \texttt{if -g} excels in delay reduction, achieving an average of 31\% logic level reduction, outperforming all other methods in this metric. Third, \texttt{orchestrate}, despite being a recent method, performs the worst, indicating limitations in its generalization ability for the various benchmarks. 

\subsection{AIG Technology Mapping} 
\begin{figure}[!t]
    \centering
    \includegraphics[width=0.9\linewidth]{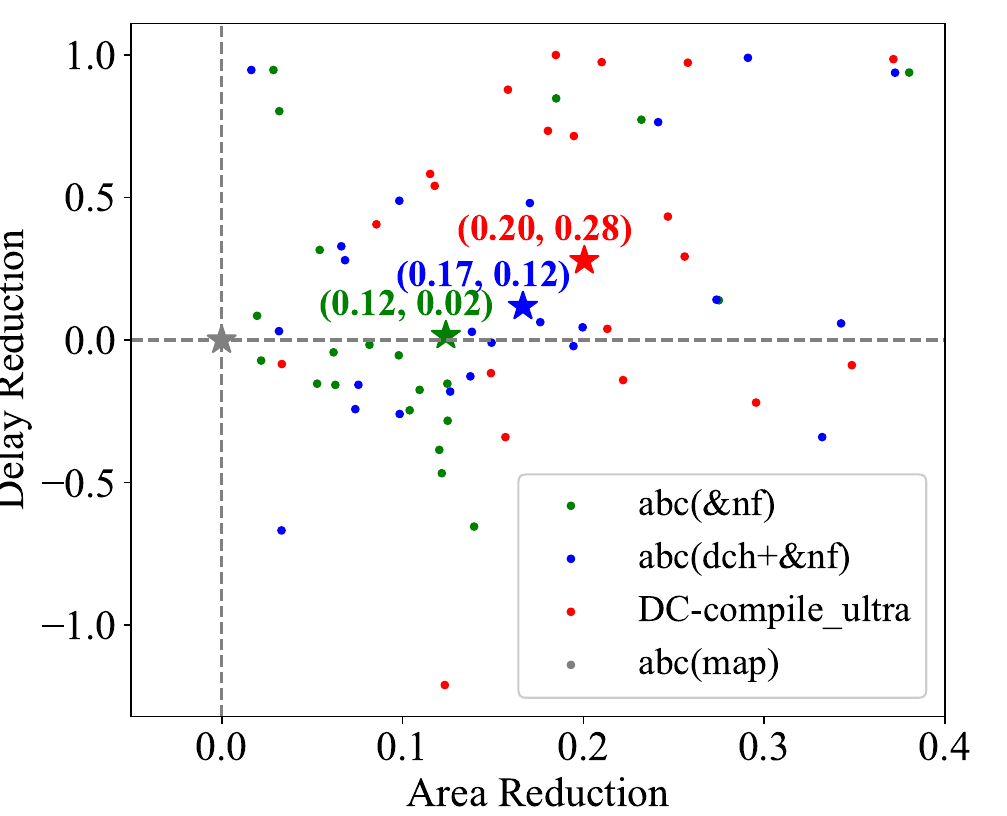}
    \vspace{-5pt}
    \caption{AIG Technology Mapping Results}
    \label{fig:aigMapping}
    \vspace{-10pt}
\end{figure}

We used the optimized AIG as input to generate mapped netlists and evaluated the mapping efficiency based on netlist area and delay. As illustrated in Fig.~\ref{fig:aigMapping}, the baseline approach is \texttt{abc(map)}. 

To assess mapping performance, we compared the ABC tool with DCU, using the ABC \texttt{map} command as a baseline. Additionally, we evaluated SOTA mapping commands such as \texttt{\&nf}, \texttt{choice}~\cite{10FPGA_mishchenko_global} within ABC.
The results indicate that, despite achieving similar performance in a few benchmarks, ABC shows a significant performance gap compared to commercial tools like DCU in most cases. 
On average, the mapped netlists produced by ABC are approximately 10\% worse in terms of area and delay, highlighting the challenges academic tools face in closing the gap with commercial solutions.

\section{AI for EDA Tasks} \label{Sec:AIApp}
The application of deep learning techniques in EDA has emerged as an attractive research direction, garnering significant interest from both the AI and EDA communities~\cite{huang2021machine, chen2024large}. Our ForgeEDA serves a dual purpose: as a dataset for training AI4EDA models and as a benchmark for evaluating their performance. We perform a series of representative tasks to train existing circuit learning models and assess their effectiveness, including probability prediction \cite{shi2023deepgate2, liu2024polargate} and equivalent gate identification \cite{shi2023deepgate2, liu2024polargate, wang2022functionality} in our experiments. 

\subsection{Probability Prediction}
\subsubsection{Task Statement} 
Probability Prediction involves determining the functionality of circuits by predicting the logic-1 probability of each gate under random simulation. This task plays a crucial role in analyzing circuit reliability and testability under probabilistic scenarios. As such, it is regarded as one of the most representative pre-training tasks in circuit learning models~\cite{shi2023deepgate2, liu2024polargate}. 

\subsubsection{Experiment Setting} 
Given an AIG graph $G$, we encode $G$ using Graph Neural Network (GNN) models as follows:
\begin{equation}
    [h_1,h_2,\ldots,h_n] = GNN(G),
\end{equation}
where $h_i$ represents the embedding of gate $i$.

Next, for a specific gate $k$, we predict its logic-1 probability using a 3-layer Multi-Layer Perceptron (MLP):
\begin{equation}
    L_{gate}^{prob} = \mathrm{MAE}(p_k, \mathrm{MLP}_{prob}(h_k)),
\end{equation}
where MAE denotes the mean absolute error. The task performance is evaluated across various GNN models: GCN\cite{GCN}, GAT\cite{GAT}, GraphSAGE\cite{GraphSAGE}, DeepGate2\cite{shi2023deepgate2}, and PolarGate\cite{liu2024polargate}.

We further divide our training dataset into 10\% and 1\% subsets to evaluate the data scalability of the models and assess the effectiveness of our expanded dataset.

\subsubsection{Result Analysis} 
\begin{table}[!t]
    \centering
    \caption{Comparison of loss value for probability prediction across different dataset splits}
    \begin{tabular}{lccc}
        \toprule
        \textbf{Model} & \textbf{Full Dataset} & \textbf{10\% Dataset } & \textbf{1\% Dataset} \\
        \midrule
        GCN~\cite{GCN} & 0.0683 & 0.1661 & 0.2497 \\
        GAT~\cite{GAT} & 0.1579 & 0.3470 & 0.3775 \\
        GraphSAGE~\cite{GraphSAGE} & 0.0307 & 0.0596 & 0.2199 \\
        DeepGate2~\cite{shi2023deepgate2} & 0.0335 & 0.0466 & 0.1965 \\
        PolarGate~\cite{liu2024polargate} & 0.0148 & 0.0164 & 0.0192 \\
        \bottomrule
    \end{tabular}
    \label{tab:prob_prediction}
    \vspace{-5pt}
\end{table}

Table~\ref{tab:prob_prediction} summarizes the performance results, where lower MAE indicates better performance. First, we observe that PolarGate achieves the best performance with an MAE of 0.0148, significantly outperforming other models. Second, as the dataset size increases, the MAE of all models decreases further. For instance, the GraphSAGE model achieves an MAE of 0.2199 on the 1\% dataset, 0.0596 on the 10\% dataset, and 0.0307 on the full dataset. These results demonstrate that ForgeEDA provides substantial benefits to AI4EDA solutions by offering a large-scale training dataset.

\subsection{Equivalent Gate Identification}
\subsubsection{Task Statement} 
Equivalent gate identification focuses on determining the functional similarity between two gates based on their truth tables~\cite{shi2023deepgate2}. This task is pivotal for applications such as Logic Equivalence Checking (LEC), SAT solving, and optimization of logic synthesis workflows.

\subsubsection{Experiment Setting} 
Given an AIG graph $G$, we encode $G$ using GNN models:
\begin{equation}
    [h_1,h_2,\ldots,h_n] = GNN(G),
\end{equation}
where $h_i$ represents the embedding of gate $i$.

For a given pair of gates $(i,j)$, their functional similarity is predicted using cosine similarity between their embeddings:
\begin{equation}
    L_{gate}^{tt\_pair} = \mathrm{MAE}(D^{gate\_tt}_{(i, j)}, \mathrm{cossim}(h_i, h_j)),
\end{equation}
where $D^{gate\_tt}_{(i, j)}$ denotes the ground-truth functional similarity derived from the truth table, and $\mathrm{cossim}(h_i, h_j)$ computes the cosine similarity. The task performance is evaluated using MAE across models: GCN\cite{GCN}, GAT\cite{GAT}, GraphSAGE\cite{GraphSAGE}, DeepGate2\cite{shi2023deepgate2}, and PolarGate\cite{liu2024polargate}. It should be noted that we do not train and evaluate FGNN~\cite{wang2022functionality}, as it predicts a hard label (equivalent or not) rather than the similarity between two logic gates. 

\subsubsection{Result Analysis} 
% \begin{table}[h]

%     \centering
%     \caption{Performance comparison for Equivalent Gate Identification task.}
%     \begin{tabular}{lc}
%         \toprule
%         \textbf{Model} & \textbf{MAE} \\
%         \midrule
%         GCN & 0.4285 \\
%         GAT & 0.4266 \\
%         GraphSAGE & 0.2789 \\
%         DeepGate2 & [Add Value] \\
%         PolarGate & [Add Value] \\
%         \bottomrule
%     \end{tabular}
    
%     \label{tab:eq_gate_ident}
% \end{table}

% \begin{table}[h]
%     \centering
%     \caption{10\% dataset ttsim}
%     \begin{tabular}{lc}
%         \toprule
%         \textbf{Model} & \textbf{MAE} \\
%         \midrule
%         GCN &  0.4396\\
%         GAT & 0.4358  \\
%         GraphSAGE &  0.4400\\
%         DeepGate2 & [Add Value] \\
%         PolarGate & [Add Value] \\
%         \bottomrule
%     \end{tabular}
% \end{table}

% \begin{table}[h]
%     \centering
%     \caption{1\% dataset ttsim}
%     \begin{tabular}{lc}
%         \toprule
%         \textbf{Model} & \textbf{MAE} \\
%         \midrule
%         GCN &  0.4641\\
%         GAT &  0.4729\\
%         GraphSAGE &  0.5815\\
%         DeepGate2 & [Add Value] \\
%         PolarGate & [Add Value] \\
%         \bottomrule
%     \end{tabular}
% \end{table}

\begin{table}[!t]
    \centering
    \caption{Comparison of loss value for equivalent gate identification task across different dataset splits}
    \begin{tabular}{lccc}
        \toprule
        \textbf{Model} & \textbf{Full Dataset} & \textbf{10\% Dataset} & \textbf{1\% Dataset } \\
        \midrule
        GCN~\cite{GCN} & 0.4285 & 0.4396 & 0.4641 \\
        GAT~\cite{GAT} & 0.4266 & 0.4358 & 0.4729 \\
        GraphSAGE~\cite{GraphSAGE} & 0.2789 & 0.4400 & 0.5815 \\
        DeepGate2~\cite{shi2023deepgate2} & 0.0813 & 0.1528 & 0.2781 \\
        PolarGate~\cite{liu2024polargate} & 0.0759 & 0.1032 & 0.1320 \\
        \bottomrule
    \end{tabular}
    \label{tab:eq_gate_ident}
    \vspace{-5pt}
\end{table}

Table \ref{tab:eq_gate_ident} presents the performance comparison (lower MAE is better). 
The results demonstrate that domain-specific models, such as DeepGate2~\cite{shi2023deepgate2} and PolarGate~\cite{liu2024polargate}, significantly outperform general-purpose GNNs. For instance, PolarGate achieves an MAE of 0.0759, compared to the much higher MAE of 0.4285 observed for GCN. Similarly, increasing the scale of the training data leads to further reductions in MAE across all models, highlighting the importance of dataset size.

\section{Conclusion}\label{Sec:Conclusion}
In this work, we present ForgeEDA, a comprehensive multimodal dataset designed to advance research and development in EDA. The dataset comprises 6 categories, 20 divisions, and 1,189 code repositories of real-world circuit designs. ForgeEDA includes multiple circuit formats, such as post-mapping netlists generated by logic synthesis tools, placed netlists produced by physical design tools, and graph representations tailored for open-source EDA tools and circuit learning models. Using ForgeEDA, we evaluate existing EDA solutions in logic synthesis and optimization, identifying significant opportunities for further advancements. Additionally, training AI4EDA models with our dataset demonstrates improved performance due to the large scale and diversity of training samples provided. We hope this work accelerates the transition of EDA towards open-source development and supports the creation of next-generation solutions for the community.

% PKU
% Beijing Natural Science Foundation (Grant No. L243001)
% National Natural Science Foundation of China（Grant No. 62032001)
% 111 Project (B18001)

% BUPT 
% This work is supported in part by the National Natural Science Foundation of China (No. U20B2045, U1936220, 62192784，62172052, 62002029，61772082).

\balance

% \section*{Acknowledgments}

\bibliographystyle{IEEEtran}
\bibliography{reference}

\end{document}